\documentstyle[12pt,epsfig]{article}

\begin{document}
\begin{titlepage}
\begin{center}

{\Large\bf Target Mass Effects in Polarized\\[2mm]
Deep Inelastic Scattering}

\end{center}
\vskip 2cm
\begin{center}
{\bf Aleksander V. Sidorov}\\
{\it Bogoliubov Theoretical Laboratory\\
Joint Institute for Nuclear Research, 141980 Dubna, Russia }
\vskip 0.5cm
{\bf Dimiter B. Stamenov \\
{\it Institute for Nuclear Research and Nuclear Energy\\
Bulgarian Academy of Sciences\\
Blvd. Tsarigradsko Chaussee 72, Sofia 1784, Bulgaria }}\\
\end{center}

\vskip 3.0cm
\begin{abstract}
\hskip -5mm The target mass effects in polarized DIS have been
studied. It was demonstrated that taking into account the first
order target mass corrections to $g_1$ a very good approximation
of the exact formula is achieved. It was also shown that their
magnitude in the preasymptotic DIS region is small except for $x>
0.65$, where their large effect is partially suppressed by the
large values of $Q^2$ due to the cut $W^2 > 4~GeV^2$. The
difference between the size of the target mass and higher twist
corrections is illustrated.



\end{abstract}

\end{titlepage}

\newpage
\setcounter{page}{1}

\section{Introduction}

One of the features of polarized DIS is that a lot of the present
data are in the preasymptotic region ($Q^2 \sim 1-5~GeV^2, 4~GeV^2
< W^2 < 10~GeV^2$). While in the unpolarized case we can cut the
low $Q^2$ and $W^2$ data in order to minimize the less known
higher twist effects, it is impossible to perform such a procedure
for the present data on the spin-dependent structure functions
without losing too much information. This is especially the case
for the HERMES, SLAC and Jefferson Lab experiments. So, to
confront correctly the QCD predictions with the experimental data
and to determine the {\it polarized} parton densities special
attention must be paid to higher twist (powers in $1/Q^2$)
corrections to the nucleon structure functions. The latter are
non-perturbative effects and cannot be calculated without using
models. That is why a {\it model independent} extraction of the
dynamical higher twists from the experimental data is important
not only for a better determination of the polarized parton
densities but also because it would lead to interesting tests of
the non-perturbative QCD regime and, in particular, of the
quark-hadron duality.

Before one can properly obtain information on the higher twist
contribution, it is important to take into account in the analysis
the so-called target mass corrections (TMCs) arising from purely
kinematic effects associated with finite values of the quantity
$4M^2x^2/Q^2$. These are also powers in $1/Q^2$ corrections,
however formally related to the twist-two operators and therefore,
unlike the higher twist ones, can be calculated without using
models. In this note we present numerical results which illustrate
the main features of the TMCs to the spin structure function $g_1$
valid in the preasymptotic DIS region. We consider that their
knowledge is useful and important in the QCD analyses of the
present and future data on polarized DIS at moderate energies.

\section{Target Mass Corrections}

\def\thefootnote{\dagger}
We will follow the method proposed by Georgi and Politzer
\cite{GP} in the case of unpolarized structure functions\footnote
{About the correspondence of this method with the Nachtmann
approach \cite{Nachtmann} see the recent paper \cite{StMel} and
the references therein.}. According to this method the target mass
corrections to the spin structure function $g_1$ have the
following form \cite{TMC_PicRid, TMC_BT}
\begin{equation}
g_1^{\rm TMC}(x, Q^2)= {1\over 2\pi i}\int _{-i\infty}^{+i\infty}
dnx^{-n} \sum _{j=0}^{\infty} \left ({M^2\over Q^2}\right )^j
{n(n+j)!\over {j!(n-1)!(n+2j)^2}} M_{n+2j}(Q^2;M=0), \label{g1TMC}
\end{equation}
where $M_n(Q^2;M=0)$ are the Cornwall-Norton (CN) moments of $g_1$
\begin{equation}
M_n(Q^2;M=0)= \int _{0}^{1} dxx^{n-1}g_1(x,Q^2; M=0) \label{CNmom}
\end{equation}
calculated in the perturbative QCD when the mass terms ${\cal
O}(M^n/Q^n)$ are neglected, or equivalently, the nucleon mass $M$
is putting equal to zero, $M=0$. As seen from (\ref{g1TMC}) TMCs
are expressed by the higher moments of $g_1(x,Q^2; M=0)$. If one
performs the infinite sum over $j$ in (\ref{g1TMC}) in order to
take into account all TMCs, $g_1$ can be written in the following
form \cite{TMC_BT}
\begin{eqnarray}
\nonumber g_1^{\rm TMC}(x,Q^2)&=&{xg_1(\xi,Q^2;M=0)\over \xi(1+
4M^2x^2/Q^2)^{3/2}}\\[5mm]
\nonumber &+&{4M^2x^2\over Q^2}~{(x+\xi)\over \xi(1+
4M^2x^2/Q^2)^2}\int _{\xi}^{1} {d\xi^{'}\over \xi^{'}}
g_1(\xi^{'},Q^2;M=0)\\[5mm]
&-&{4M^2x^2\over Q^2}~{(2 - 4M^2x^2/Q^2)\over 2(1+
4M^2x^2/Q^2)^{5/2}}\int _{\xi}^{1} {d\xi^{'}\over \xi^{'}} \int
_{\xi^{'}}^{1} {d\xi^{''}\over \xi^{''}}g_1(\xi^{''},Q^2;M=0),~~~~
\label{g1TMCtot}
\end{eqnarray}
where $g_1(x,Q^2;M=0)$ is the well known pQCD expression for g1
(logarithmic in $Q^2$) obtained in LO or NLO approximation {\it
neglecting} the target mass corrections, and
\begin{equation}
\xi= {2x\over {1+ (1+4M^2x^2/Q^2)^{1/2}}} \label{Nachtvariable}
\end{equation}
is the Nachtmann variable. In Eqs. (\ref{g1TMC}-\ref{g1TMCtot}) we
have dropped the nucleon target label N.

Let us now discuss the numerical results on the target mass
corrections to the spin structure function $g_1$. In our analysis
we will mainly concentrate on their features in the preasymptotic
DIS region (the invariant mass $W > 2~GeV$ and moderate values of
$Q^2:1 - 10~ GeV^2$). In the calculations of the target mass
effects we have used for $g_1(x,Q^2;M=0)$ our recent NLO results
from $'~g_1/F_1~'$ fit presented in \cite{LSS05pos}. For the
further discussion it is useful to rewrite (\ref{g1TMC}) in the
form:
\begin{eqnarray}
\nonumber g_1^{\rm TMC}(x, Q^2)&=&g_1(x, Q^2;M=0) + {M^2\over
Q^2}g_1^{(1)\rm TMC}(x, Q^2)\\
&+& \left ({M^2\over Q^2}\right )^2 g_1^{(2)\rm TMC}(x, Q^2) +
\left ({M^2\over Q^2}\right )^3g_1^{(3)\rm TMC}(x, Q^2) + {\cal
O}(M^8/Q^8), \label{g1TMC1}
\end{eqnarray}
where
\begin{equation}
g_1^{(j)\rm TMC}(x, Q^2)={1\over 2\pi i}\int
_{-i\infty}^{+i\infty} dnx^{-n}{n(n+j)!\over {j!(n-1)!(n+2j)^2}}
M_{n+2j}(Q^2;M=0),~ j=1, 2,... \label{g1fixedorder}
\end{equation}

The questions we address here are {\it i)} How large are the
target mass effects in the preasymptotic DIS region and {\it ii)}
How fast the series (\ref{g1TMC}) converges, or in other words how
large is the difference between $g_1^{\rm TMC}$ calculated by
(\ref{g1TMCtot}) and $g_1^{\rm TMC,Q^2}$ obtained when the TMCs
are calculated only up to the first order in $(M^2/Q^2)$ (the
first two terms in (\ref{g1TMC}) or equivalently, in
(\ref{g1TMC1})). We will focus on the proton target, but in the
end of our discussion we will mention also our results concerning
the neutron spin structure function $g_1^n(x,Q^2)$. Note that it
is of interest from a theoretical point of view to examine the
convergence of the TMCs irrespective to the size of the dynamical
HT. For practical purposes, however, there is no sense to use the
exact formula (\ref{g1TMCtot}) in the analysis of the experimental
data if the dynamical HT are not negligible to those of the TMCs.
In this case the target mass and dynamical higher twist
corrections have to be taken into account up to the same finite
order in ${\cal O}(1/Q^2)$.

The target mass corrections at fixed order in $(M^2/Q^2)$ in the
expansion (\ref{g1TMC1}) are illustrated in Fig. 1 where the
notations
\begin{equation}
\tilde {g_1}^{(j)\rm TMC}(x,Q^2) = \left ({M^2\over Q^2}\right
)^jg_1^{(j)\rm TMC}(x,Q^2),~~ j=1,2... \label{notation}
\end{equation}
are used. The TMCs are given outside the resonance region, {\it
i.e.}, $W^2> 4~{\rm GeV^2}$, where $W^2=M^2+Q^2(1-x)/x$, at
different fixed values of $Q^2:Q^2=1~{\rm GeV^2}$ (black curves),
$3~{\rm GeV^2}$ (red curves) and $10~{\rm GeV^2}$ (green curves).
One can see from Fig. 1 that the series (\ref{g1TMC}), or
equivalently (\ref{g1TMC1}), is an alternating series and the size
of TMCs of higher then the first order in $(M^2/Q^2)$ are small,
especially the target mass corrections of third order in
$(M^2/Q^2)$, $\tilde {g_1}^{(3)\rm TMC}(x,Q^2)$, which are smaller
than
$6.10^{-4}(Q^2=1~GeV^2);~2,5.10^{-4}(Q^2=3~GeV^2);~5.10^{-6}(Q^2=10~GeV^2)$.
\begin{figure}[bht]
\centerline{ \epsfxsize=3.8in\epsfbox{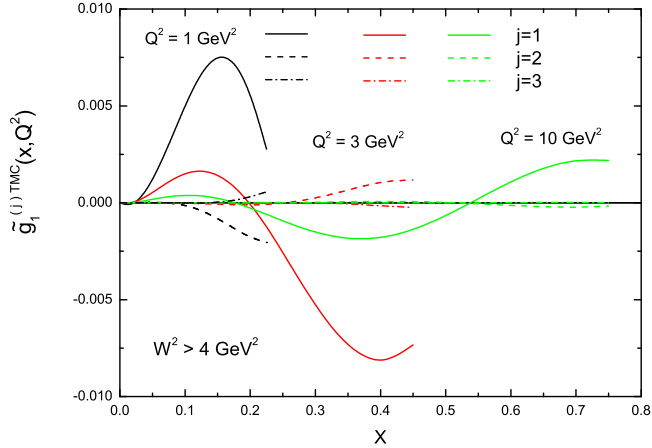} } \caption{Target
mass corrections $\tilde {g}_{1, p}^{(j){\rm TMC}}(x,Q^2)$ at
fixed orders in $(M^2/Q^2)^j$ for three different values of $Q^2$
(see the text).}
\end{figure}

In order to estimate the target mass effects in the preasymptotic
DIS region we demonstrate also the relative changes of $g_1^p$
taking into account TMCs up to order ${\cal O}(M^2/Q^2)$, ${\cal
O}(M^6/Q^6)$ and all orders of $M^2/Q^2$ (Eq. \ref{g1TMCtot}).
These effects are presented for $Q^2=1~GeV^2$ (Fig. 2),
$Q^2=3~GeV^2$ and $Q^2=10~GeV^2$ (Fig. 3), respectively. The
notations used in these figures are as follows:
\begin{eqnarray}
\nonumber g_1^{{\rm TMC},Q^2}&=&g_1(M=0) + {M^2\over Q^2}
g_1^{(1)\rm TMC},\\[3mm]
\nonumber g_1^{{\rm TMC},Q^6}&=&g_1(M=0) + \sum _{j=1}^{3} \left
({M^2\over Q^2}\right )^jg_1^{(j)\rm TMC},\\[3mm]
g_1^{\rm TMC,tot}&=&g_1^{\rm TMC}({\rm Eq.~ 3}). \label{g1Q2Q6}
\end{eqnarray}

Note that to calculate $g_1^{\rm TMC,tot}(x,Q^2)$ we have used the
prescription given in \cite{TMC_BT} that $g_1(\xi,Q^2;M=0)$ in the
RHS of Eq. \ref{g1TMCtot} vanishes for $\xi > \xi_0$, where
$\xi_0=\xi(x=1)$.
\begin{figure}[bht]
\centerline{ \epsfxsize=3.3in\epsfbox{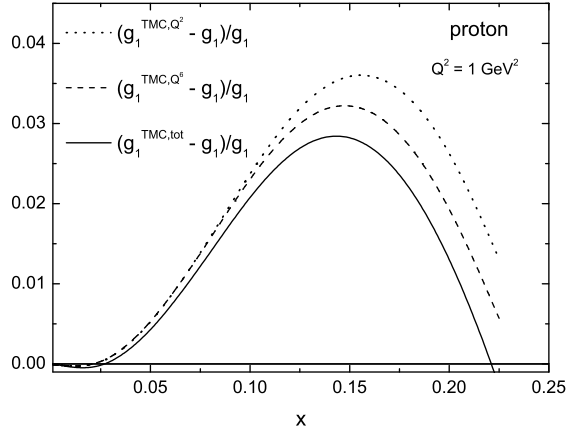} } \caption{The
proton target mass effects at $Q^2=1~{\rm GeV^2}$ (see the text).
}
\end{figure}

\def\thefootnote{\dagger}
In Figures 2 and 3 the dot, dashed and solid curves correspond to
the change of $g_1$ proton when the first, third and all orders in
$(M^2/Q^2)$, respectively, are taken into account. As seen from
these figures, the sign of the target mass corrections depends on
the $x$ region and the maximum change of the magnitude of $g_1$
due to the first order TMCs is 3.6\% at $Q^2=1~{\rm GeV^2}$, 6\%
at $Q^2=3~{\rm GeV^2}$ and 28.6\% at $Q^2=10~{\rm
GeV^2}$\footnote{The fact that in the polarized case the target
mass corrections at first order in $M^2/Q^2$ are generally small
was first established in \cite{TMC_PicRid}.}. The TM effects are
large in the large $x$ and small $Q^2$ region. However, because of
the cut $W^2> 4~{\rm GeV^2}$, the large $x$ region at small $Q^2$
is outside of the preasymptotic DIS region and their effects are
much smaller than those in the resonance one. One can see also
from Figs. 2 and 3 that taking into account a first order TMCs to
$g_1$ a good approximation of the exact equation (\ref{g1TMC}) is
already achieved. The deviation of this approximation of $g_1$
from $g_1^{\rm TMC,tot}(x,Q^2)$, is not more than 1.5\% for
$Q^2=1~{\rm GeV^2}$, 0.8\% for $Q^2=3~{\rm GeV^2}$ and 1.8\% for
$Q^2=10~{\rm GeV^2}$. The dashed curves, which correspond to
$g_1^{{\rm TMC},Q^6}$ (the TMCs are taken into account up to a
third order in $M^2/Q^2$), practically coincide with $\Delta
g_1^{\rm TMC,tot}/g_1$ (solid curves), especially in the cases of
$Q^2=3~{\rm GeV^2}$ and $Q^2=10~{\rm GeV^2}$. This observation is
a consequence of the fact we have already discussed above that the
series of the target mass corrections (\ref{g1TMC}) is an
alternative sign one.
\begin{figure}[bht]
\centerline{ \epsfxsize=3.0in\epsfbox{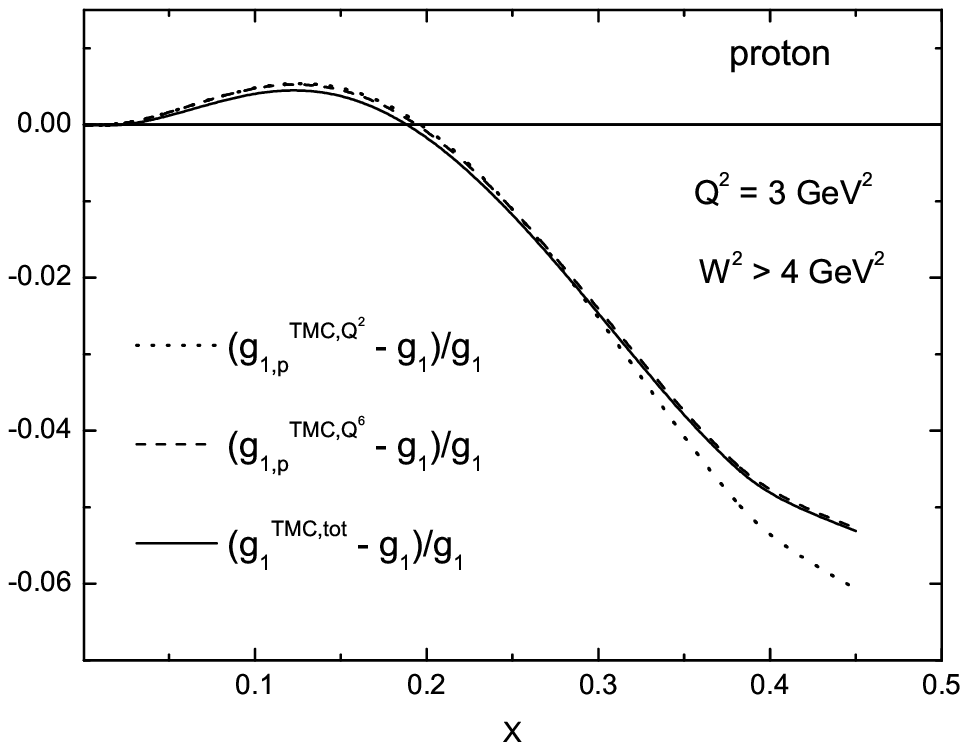}
\epsfxsize=3.0in\epsfbox{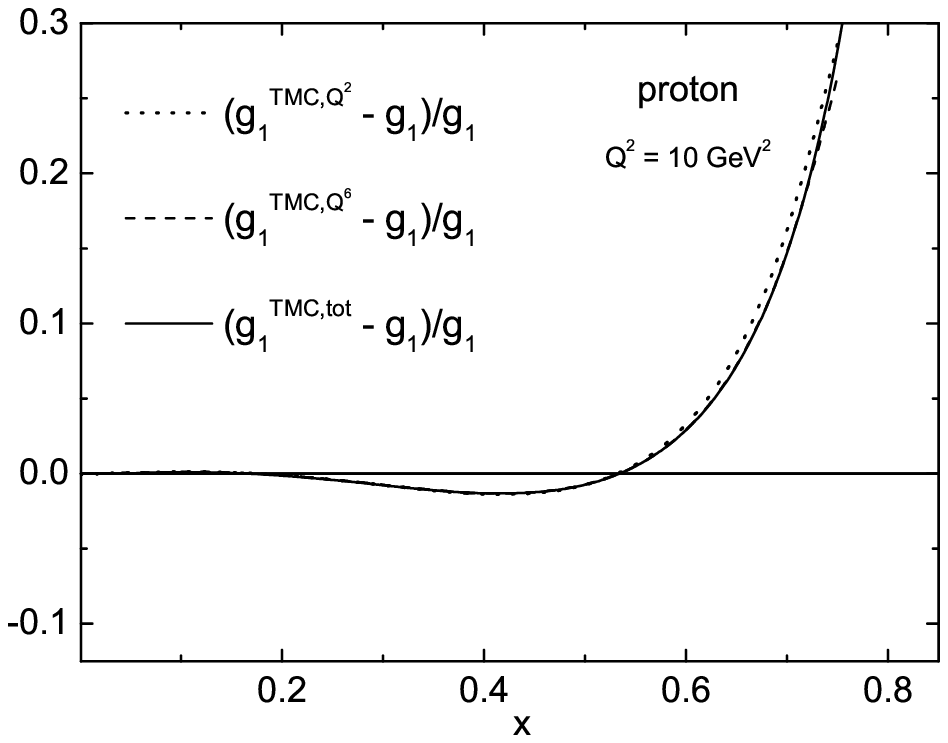}} \caption{The proton target
mass effects at $Q^2=3,~10~{\rm GeV^2}$ (see the text). }
\end{figure}

In Fig. 4 we compare the TMCs to neutron spin structure function
$g_1^n(x,Q^2)$ with those to $g_1^p$. As seen from Fig. 4 its
magnitude is much smaller than that one in the proton case. This
is due to the fact that the negative $g_1^n$ crosses zero at
$x\approx 0.4$ and becomes positive at $x$ higher than 0.4
\cite{JLab}. As a result, the values of the moments of $g_1^n$ in
the RHS of (\ref{g1TMC}) are much smaller than those of $g_1^p$.
The main conclusion about the proton TMCs, namely, that the first
order TMCs is a good approximation of their total account, holds
for the neutron target too.
\begin{figure}[bht]
\centerline{ \epsfxsize=3.2in\epsfbox{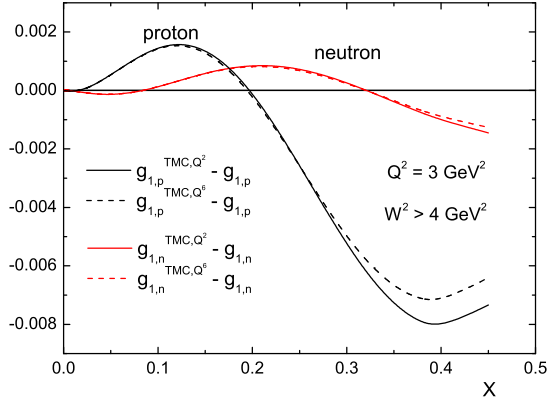} }
\caption{Comparison between the target mass corrections to $g_1^n$
(red curves) and $g_1^p$ (black curves) at $Q^2=3~{\rm GeV^2}$.  }
\end{figure}

Finally, we would like to compare the size of the TMCs with that
of the higher twist power corrections to the spin structure
function $g_1$. Taking into account only the first order
corrections in $1/Q^2$, $g_1$ has the following form:
\begin{equation}
g_1(x, Q^2)=g_1(x, Q^2;M=0) + {M^2\over Q^2}g_1^{(1)\rm TMC}(x,
Q^2) + {h(x,Q^2)\over Q^2} + {\cal O}(1/Q^4), \label{g1Q2}
\end{equation}
where $h(x,Q^2)$ are the dynamical higher twist ($\tau=3$ and
$\tau=4$) corrections to $g_1$, which are related to multi-parton
correlations in the nucleon. The latter are non-perturbative
effects and cannot be calculated without using models.

In Fig. 5 we compare the first order TMCs, $\tilde {g_1}^{(1)\rm
TMC}(x,Q^2)$~(see Eq. \ref{notation}), with the values of the HT
corrections, ${h(x,Q^2)/Q^2}$, extracted in a model independent
way from the world data on polarized DIS \cite{LSS05pos}. One can
see from Fig. 5 that: {\it i)} In the neutron case the target mass
contribution in $g_1$ is negligible compared to the higher twist
effects and {\it ii)} In the proton case the TMCs are essential at
$x>0.3$ while for $x < 0.3$ their magnitude is much smaller than
that of the HT corrections. Note also the different shapes of TM
and HT corrections.
\begin{figure}[bht]
\centerline{ \epsfxsize=3.0in\epsfbox{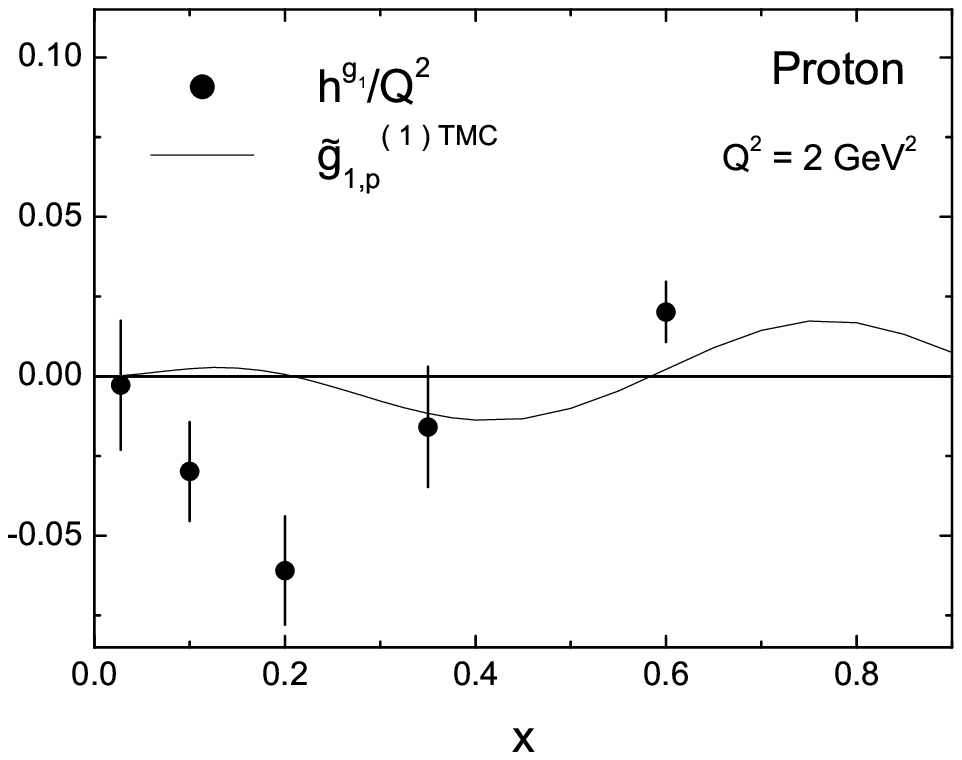}
\epsfxsize=3.0in\epsfbox{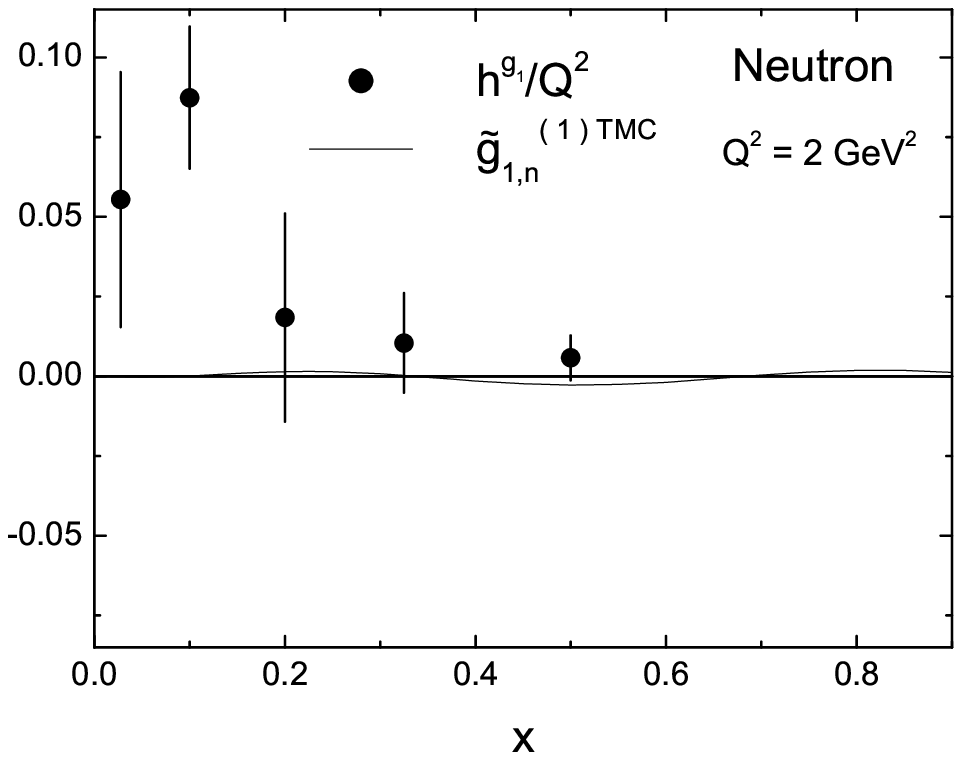}} \caption{Comparison between
the first order target mass corrections to $g_1$ (solid curves)
and the HT (twist-3 and 4) corrections at $Q^2=2~{\rm GeV^2}$. }
\end{figure}
As a result of this analysis we conclude that in the preasymptotic
DIS region the higher twist contribution in $g_1$ becomes more and
more important, so the TMCs cannot be trusted alone.

\section{Summary}
We have studied the target mass effects in polarized DIS
scattering which would be important in preasymptotic region. It
was demonstrated that accounting for the first order target mass
corrections to $g_1$ a very good approximation of the exact
formula (\ref{g1TMCtot}) is achieved. It was also shown that the
size of the TMCs in the preasymptotic DIS region is small ($<
6\%$) except for $x> 0.65$ where their effect is a partially
suppressed by the large values of $Q^2$ due to the cut $W^2 >
4~GeV^2$. Compared to the higher twist effects the contribution of
target mass corrections to the spin structure function $g_1$ in
the preasymptotic DIS region is insignificant except for $x>0.3$
in the proton case. Nevertheless, to extract correctly the unknown
high twist effects from the experimental data, the calculable
target mass corrections to the spin structure functions should be
taken into account.

\vskip 6mm { \bf Acknowledgments} \vskip 4mm We would like to
thank E. Leader for useful discussions. This research was
supported by the UK Royal Society and the JINR-Bulgaria
Collaborative Grants, and by the RFBR (No 05-01-00992,
05-02-17748, 06-02-16215). \\


\end{document}